# Floquet-Magnus Expansion and Fer Expansion Approaches Revisited to Investigate the Chemical Shift Anisotropy During the Triple Oscillating Field Technique Irradiation in Solid-State NMR

Eugene Stephane Mananga[1,2,3*]

[1*]*PH. D Programs of Physics & Chemistry, Graduate Center, The City University of New York, New York 10016, USA*

[2*]*Department of Engineering, Physics, and Technology, Bronx Community College, The City University of New York, 2155 University Avenue, CPH 118, Bronx, New York 10453, USA*

[3*]*Department of Applied Physics, New York University, 6 Metro-tech Center, Brooklyn, New York 11201, USA*

**Abstract**

The Floquet-Magnus and Fer expansion schemes were introduced in solid-state nuclear magnetic resonance (NMR) in 2011 and 2006, respectively. Key features of the Floquet-Magnus expansion are its ability to account for the calculations developed in a finite-dimensional Hilbert space instead of an infinite-dimensional space within the Floquet theory as well as its use of its distinguishable function, $\Lambda_n(t), n = 1, 2, 3, ...$, not available in other concurrent theories such as average Hamiltonian theory, Floquet theory, and Fer expansion. The distinguishable function facilitates the evaluation of the spin behavior in between the stroboscopic observation points. This article focuses on revisiting the Floquet-Magnus and Fer expansion approaches, and applying both methods to calculate the effective Hamiltonians and propagators, which control the spin system evolution during the Triple Oscillating Field Technique radiation experiment (TOFU). The TOFU pulse sequence is an important technique that was shown to avoid the dipolar truncation problem and form a new basis for accurate distance measurement by solid-state NMR. We take advantage of the interaction frequencies and the time modulation arising from the TOFU pulse sequence allowing selective recoupling of specific terms in the Hamiltonian that fulfill determined specific conditions. The work presented here unifies and generalizes results of the Floquet-Magnus and Fer expansions, and delivers illustrations of novel springs that boost previous applications that are based on the classical information. We believe that the revisited approaches of this work and the derived expressions can serve as useful information and numerical tools for time evolution in time-resolved spectroscopy, quantum control, and open system quantum dynamics.

## I. Introduction

Over the past two decades, solid-state nuclear magnetic resonance (NMR) spectroscopy has revealed its ability of producing atomic-resolution structures of solids and semi-solids[1]. This has opened new possibilities to elucidate molecular structure and dynamics in systems, which in many cases cannot be obtained by alternative ways[2]. Most solid-state NMR experiments are established on a combination of magic-angle spinning (MAS) and dipolar recoupling[3]. This technique can also be combined with cross polarization to increase the spectral sensitivity of rare and low-gamma nuclei such as $^{13}$C, $^{15}$N in biopolymers or other organic solids[2]. Therefore, MAS NMR techniques have improved to determine complete structure of systems[4-6]. Methods to describe the time evolution of spin systems under time-dependent Hamiltonians play an important role among the tools used to analyze solid-state nuclear magnetic resonance experiments. The technique of spin recoupling, first proposed by Andrew[7], has evolved into a universal and essential part of high-resolution NMR spectroscopy. The general idea of spin recoupling is that the radio-frequency irradiation is used to manipulate the spin part of the Hamiltonian to reintroduce certain terms in the Hamiltonian[8-10]. In many situations, the resulting time dependent Hamiltonian becomes periodic with possibly multiple incommensurate frequencies. MAS NMR experiments are preferred for structural studies of biomolecular systems that exhibit low solubility or lack long range order and therefore cannot be addressed with the traditional tools of structural biology, solution NMR or X-ray diffraction[11]. The basic approach consisting of reintroducing dipolar couplings with a train of pulses at the 13C and 15N frequencies[5,12] while simultaneously decoupling the 1H spins from the 15N-13C spin dynamics was employed in the development of the majority of heteronuclear recoupling sequences including rotational echo double resonance (REDOR)[13], transferred echo double resonance (TEDOR)[14], frequency selective (FS)-REDOR[15], Z-filtered (ZF-) and band-selective (BASE-) TEDOR, and frequency selective (FS-) TEDOR[16], which enable accurate N-C distance measurements, and have been especially important in the determination of high resolution 3D structures[17]. MAS NMR experiments average 2nd rank tensor interactions such as the chemical shift anisotropy and dipolar interactions, and therefore yield high resolution spectra. One of the most significant challenges encountered in the recoupling pulse sequences is the requirement to accommodate high power rf irradiation on all three channels (1H, 13C, and 15N) during the mixing periods. Nearly a decade and half ago, Khaneja and Nielsen[18] proposed a new concept for homonuclear dipolar recoupling in MAS solid-state NMR experiments, which avoids the problem of dipolar truncation. With the introduction of the triple oscillating field technique (TOFU), Khaneja and co-worker demonstrated that the TOFU is an efficient means to accomplish broadband dipolar recoupling of homonuclear spins, while decoupling heteronuclear dipolar couplings and anisotropic chemical shifts and retaining influence from isotropic chemical shifts[18].

* Corresponding author. Telephone +1 646 345 4613; Fax: +1 718-289-6403
E-mail addresses: esm041@mail.harvard.edu or emananga@gradcenter.cuny.edu (Eugene S. Mananga)

In this article, we applied the two developing approaches which describe the behavior or spin dynamics in solid-state NMR, namely the Floquet-Magnus expansion (FME)[2,19-33] and the Fer expansion (FE)[34-48] to control the spin dynamics on the TOFU radiation experiment. We investigated the orders to which the FME and the FE approaches are equivalent or different for the chemical shielding Hamiltonian during the application of the TOFU pulse sequence radiation experiment. The TOFU experiment allows for broadband recoupling of Ising interaction Hamiltonian, $2I_zS_z$, while simultaneously ensuring that the chemical shifts of the spins are maintained and the heteronuclear coupling are decoupling[18,49]. This method circumvents the dipolar truncation problem and form a new basis for accurate distance measurement par solid-state NMR[50-55].

The Floquet-Magnus expansion[2] and the Fer expansion[36] schemes were introduced in solid-state NMR in 2011 and 2006, respectively. Key features of the Floquet magnus expansion are its ability to account for the calculations developed in a finite-dimensional Hilbert space instead of an infinite-dimensional space within the Floquet theory as well as its use of its distinguishable function, $\Lambda_n(t), n = 1, 2, 3, ...$, from other theories such as average Hamiltonian theory, Floquet theory, and Fer expansion, which facilitates the evaluation of the spin behavior in between the stroboscopic observation points. Notably, the $\Lambda_n(t)$ functions represents the $n^{th}$ – order term of the argument of the operator that introduces the frame such that the spin system operator is varying under the time-independent Hamiltonian, F. This view presents the functions $\Lambda_n(t)$ as the argument of the operator that introduces the frame that varies under the time-independent Hamiltonian, F. Such theoretical advance has been facilitated by the development of a more general representation of the evolution operator, which removes the constraint of a stroboscopic observation. This current article focuses on applying the Floquet-Magnus and Fer expansion approaches for the calculation of effective Hamiltonians and propagators to control the spin system evolution during the Triple Oscillating Field Technique radiation experiment. Our work unifies and generalizes existing results of the Floquet-Magnus and Fer expansions and delivers illustrations of novel springs that boost previous applications that are based on the classical information. The generality of this work points to potential applications in problems related to theoretical developments of spectroscopy as well as interdisciplinary research areas whenever they include spin dynamics concepts.

In this manuscript, we explain in detail the spin dynamics mechanism during the application of the TOFU technique. Notably, we show that analytical expressions derived using FME and FE approaches facilitates a clear understanding of the associated spin physics. The paper is organized as follows. In sections II, we revisited the TOFU Pulse Sequence, and in sections III and IV, we revisited the theoretical formalism of the FME and FE, respectively. In sections V and VI, we performed the applications of FME and FE to Chemical Shielding Hamiltonian during the TOFU Pulse Sequence Radiation. The comparison and discussion are presented in section VII, and the conclusion in section VIII. We also presented an extended appendix containing detailed derivations and calculations of the results.

## II. Revisiting the TOFU Pulse Sequence

The triple oscillating field technique called TOFU was introduced a decade and half ago by Khaneja and Nielden[18]. The TOFU pulse sequence allows for broadband recoupling of Ising interaction Hamiltonian ($2I_ZS_Z$), while simultaneously ensuring that the chemical shifts of the spins are maintained and the heteronuclear couplings are decoupled[49]. This technique was shown to avoid the dipolar truncation problem and form a new basis for accurate distance measurement by solid-state NMR. Consider two coupled homonuclear spins I and S under magic-angle-spinning (MAS) conditions. Within the high-field approximation and under conditions of MAS, the Hamiltonian of the spin system takes the form[49],

$$H_0(t) = \omega_I(t)T_{1,0}^I + \omega_S(t)T_{1,0}^S + \omega_{IS}(t)\sqrt{6}T_{2,0}^{IS}, \tag{1}$$

where the first two terms include isotropic/anisotropic chemical shifts for the two spins while the third term denotes the dipole-dipole coupling interaction. These interaction strengths may be expressed in terms of a Fourier series,

$$\omega_{\lambda,m'}(t) = \sum_{m=-2}^{2} \omega_{\lambda,m'}^{(m)} \exp(im\omega_r t) \quad (2)$$

where $\omega_r$ is the sample spinning frequency in angular units and $\lambda = I, S, IS$ is a signature of the specific interaction such as the chemical shift, dipole-dipole coupling (through this the internuclear distance), J coupling, and quadrupolar coupling[56-58]. The Fourier coefficients are

$$\omega_{\lambda,m'}^{(m)} = \omega_{iso}^{\lambda} \delta_{m,0} + \omega_{aniso}^{\lambda} \left\{ D_{0,-m}^{(2)}(\Omega_{PR}^{\lambda}) - \frac{\eta^{\lambda}}{\sqrt{6}} \left[ D_{-2,-m}^{(2)}(\Omega_{PR}^{\lambda}) + D_{2,-m}^{(2)}(\Omega_{PR}^{\lambda}) \right] \right\} d_{-m,m'}^{(2)}(\beta_{RL}), \quad (3)$$

where $\delta_{m,0}$ is a standard Kronecker delta and the constants specifying the isotropic ($\omega_{iso}^{\lambda}$) and anisotropic ($\omega_{aniso}^{\lambda}, \eta^{\lambda}$) contributions to the Fourier coefficients (for the i-spin) are given in angular frequency units by

$$\omega_{iso}^{\lambda} = \omega_0^i \delta_{iso}^i - \omega_{ref} \quad (4\text{-a})$$

$$\omega_{aniso}^{\lambda} = \omega_0^i \delta_{aniso}^i \quad (4\text{-b})$$

$$\eta^{\lambda} = \eta_{CS}^i \quad (4\text{-c})$$

$$\delta_{iso}^i = \frac{1}{3}\left(\delta_{xx}^i + \delta_{yy}^i + \delta_{zz}^i\right) \quad (4\text{-d})$$

$$\delta_{aniso}^i = \delta_{zz}^i - \delta_{iso}^i \quad (4\text{-e})$$

$$\eta_{CS}^i = \left(\delta_{yy}^i - \delta_{zz}^i\right) \Big/ \delta_{aniso}^i \quad (4\text{-f})$$

$$\omega_0^i = -\gamma_i B_0. \quad (4\text{-g})$$

$\delta_i$ is the gyromagnetic ratio and $B_0$ the flux density of the static magnetic field. $\omega_{ref}$ is an optional rotating frame reference frequency. The principal elements are ordered according to

$$\left|\delta_{zz}^i - \delta_{iso}^i\right| \geq \left|\delta_{xx}^i - \delta_{iso}^i\right| \geq \left|\delta_{yy}^i - \delta_{iso}^i\right|. \quad (4\text{-h})$$

The second-rank Wigner ($D^{(2)}$) and reduced Wigner ($d^{(2)}$) rotation matrices in the Eq. (3) reveal the orientation dependence for the anisotropic interactions. Furthermore, the Wigner rotation matrices, $D_{m',m}^{(l)}(\Omega_{AB}^{\lambda})$, describe the coordinate transformation between axis systems A and B, according to a set of three Euler angles[59,60],

$$\Omega_{AB}^{\lambda} = \{\alpha_{AB}^{\lambda}, \beta_{AB}^{\lambda}, \gamma_{AB}^{\lambda}\} \quad (4\text{-i})$$

such as,

$$D^{(l)}_{m',m}(\Omega^\lambda_{AB}) = \exp(-im'\alpha^\lambda_{AB})d^{(l)}_{m',m}(\beta^\lambda_{AB})\exp(-im\gamma^\lambda_{AB}), \quad (4\text{-j})$$

where $d^{(l)}_{m',m}(\beta^\lambda_{AB})$ is the reduced Wigner matrix. Any given interaction $\lambda$ describes these matrices coordinate transformations from the principal-axis frame ($P^\lambda$) to the laboratory-fixed frame (L), as well as a rotor-fixed frame (R) such as described by Bak et al.[56]. For simplicity, the Eq. (3) assumes that the principal axis frame coincides with the crystallite-fixed frame.

With the aim of recoupling the dipole-dipole coupling interaction while maintaining strong contributions from the isotropic chemical shift terms to truncate non-secular terms in the dipolar coupling Hamiltonian, the pioneering TOFU technique experiment[18] uses a rotor synchronized, time-dependent rf Hamiltonian takes the general form

$$H_{rf}(t) = A(t)\big(cos\phi(t)F_x + sin\phi(t)F_y\big) + \omega(t)F_z \quad (5)$$

where

$$F_q = I_q + S_q \ (q = x, y, z). \quad (7)$$

By appropriate choice of amplitude, phase, and offset, the rf field can be written as

$$H_{rf}(t) = CF_x + Cexp(-iCtF_x)F_y \exp(iCtF_x)$$
$$+ Bexp(-iCtF_x)\exp(-iCtF_y)F_z\exp(iCtF_y)\exp(iCtF_x). \quad (8)$$

This particular setting of the rf fields may be appreciated by performing a series of coordinate transformations[18]. The TOFU technique was recently improved by introducing the four-oscillating field dipolar recoupling technique in a one-dimensional (1D) setup that allowed the user to extract accurate $^{13}$C-$^{13}$C distances[49]. In this work, we applied the FME and FE to the original TOFU (Triple Oscillating Field Technique) to understand and investigate the spin dynamics during the TOFU pulse sequence radiation for the chemical shielding Hamiltonian. The overall transformation to the recoupling frame represented as

$$\tilde{A} = U^+(t)AU(t), \quad (9)$$

with

$$U(t) = \exp(-iCtF_x)\exp(-iCtF_y), \quad (10)$$

where $F_q$ is defined by the Eq. (7) and C is the magnetic field strength described by Khaneja and co-worker[18]. In this article, within the recoupling frame, we only consider the expression for the chemical shielding Hamiltonian leaving other interactions for future work and development.

## III. Revisiting the Theoretical formalism of the FME

The idea of a formulation of FME applied directly to periodically driven many-body quantum systems has attracted much attention over recent years, in particular is solid-state NMR. The FME benefit from the fusion of average Hamiltonian theory (Magnus expansion)[61-70] and Floquet theory[71-75] as well as the concise natural treatment of time-periodic Hamiltonian that both approaches (average Hamiltonian and Floquet theories) provide. The divergence of the FME anticipate interesting physical meaning[2,19,76]. Part of this article focused on the FME that gives a formal expression of the effective Hamiltonian on the system. Plausibly, the most noteworthy difference of the Floquet−Magnus expansion with other approaches such as Magnus expansion, Floquet theory, or Fer expansion may be that it expands a propagator in the form of a more basic representation of the evolution operators as[2,77]

$$U(t) = P(t)e^{-itF}P^{+}(0), \tag{11}$$

which removes the constraint of a stroboscopic observation. P(t) can be seen as the operator that introduces the frame such that the density operator is varying under the time independent Hamiltonian F. The FME is obtained by representing the solution of the time dependent Schrödinger equation

$$\frac{dU(t)}{dt} = -iHU(t) \tag{12}$$

in the form of eq. (11) and using the following exponential ansatz

$$P(t) = exp\{-i\Lambda(t)\} \tag{13}$$

where the function $\Lambda(t)$ is the argument of the operator P(t). Introducing the expansions

$$\Lambda(t) = \sum_n \Lambda_n(t), \tag{14}$$

and

$$F = \sum_n F_n, \tag{15}$$

the FME expansion can be summarized as

$$\Lambda(t) = \Lambda_n(0) + \int_0^t G_n(u)du - tF_n \tag{16}$$

where the first functions $G_n(t)$ are defined in the literature[2]. The above Eq. (16) includes two operators $\Lambda_n(t)$ and $F_n$ independent of each other. Indeed, the periodicity conditions,

$$\Lambda_n(\tau_C) = \Lambda_n(0) \tag{17}$$

where $\tau_C$ is the period of the modulation,

$$H(\tau_C + t) = H(t) \tag{18}$$

defines $F_n$ as

$$F_n = \frac{1}{\tau_C} \int_0^{\tau_C} G_n(u) du \tag{19}$$

such that we are free to choose the operators $\Lambda_n(0)$, i.e., the boundary conditions. At first glance, the choice $\Lambda_n(0) = 0$ (*i.e.*, $P(0) = 1$) appears as the simplest. As was shown in our previous work[2], in this case the FME reduces to the Magnus expansion (ME), which validity is restricted to stroboscopic observation, as was discussed in numerous papers[2,19,26,29,76]. A much better choice, without any restriction on the observation time, is given by the general rule,

$$\int_0^{\tau_C} \Lambda_n(u) du = 0 \tag{20}$$

which was shown to simplify higher order terms in the $F_n$ expansion. The FME is also known to give a formal expression of the Floquet Hamiltonian as follows

$$H_F = \sum_{n=0}^{\infty} T^n \Omega_n \tag{21}$$

where explicit forms of the terms $\{\Omega_n\}_{n=0}^{\infty}$ are given by

$$\Omega_n = \frac{1}{(n+1)^2} \sum_\sigma (-1)^{n-\theta(\sigma)} \frac{\theta(\sigma)!\,(n-\theta(\sigma))!}{n!} \frac{1}{i^n T^{n+1}}$$
$$\times \int_0^T dt_{n+1} \ldots \int_0^{t_3} dt_2 \int_0^{t_2} dt_1 \left[ H(t_{\sigma(n+1)}), [H(t_{\sigma(n)}), \ldots, [H(t_{\sigma(2)}), H(t_{\sigma(1)})] \ldots] \right]. \tag{22}$$

The parameter $\sigma$ is the permutation and $\theta(\sigma) := \sum_{i=1}^{n} \theta(\sigma(i+1) - \sigma(i))$, with $\theta(.)$ the usual step function[2,19]. The FME is useful to investigate periodically driven system when the period T of the driving is sufficiently small. The above Eq. (22) is useful especially for high-frequency limit in finite-size systems, where higher-order contribution is negligible. The FME can also be truncated up to the n$^{th}$ order, which is defined as

$$H_F^{(n)} := \sum_{m=0}^{n} T^m \Omega_m \tag{23}$$

### IV. Revisiting the Theoretical formalism of the FE

The Fer expansion was developed by Fer more than half a century ago[35] and further thrived by Klarsfeld et al.[78]. Recently, Madhu et al.[36] introduced the approach to solid-state NMR, while Takegoshi et al.[77] as well as other authors such as[34,79,80] contributed to exhaustive descriptions of Fer expansion in solid-state NMR. In this section, we recapitulate and revisted the results of Madhu[36], Takegoshi[77], and Ganguly[79,80] for the Fer expansion without going into detail. In the Fer expansion, the evolution operator is expressed in terms an infinite product of exponential operators such as,

$$U(\tau_C) = \prod_{n=0}^{\infty} exp\{-i\tau_C \bar{H}_{Fer}^{(n)}\} \tag{24}$$

Depending on the ordering of the operators, the time-propagator, $U(t)$, derived from Fer expansion is classified into two categories[36,77-80]: The Right running Fer expansion and the Left running expansion, which are represented in the following,

(a) Right running Fer expansion

$$U_R(t) = exp(\lambda F_1(t)) exp(\lambda^2 F_2(t)) exp(\lambda^3 F_3(t)) \ldots exp(\lambda^{n-1} F_{n-1}(t)) exp(\lambda^n F_n(t)) U_n(t). \tag{25}$$

Next, using the form of the $F_n(t)$ operators, we can evaluate the density operator (Right running Fer expansion) at time t as well as derive the signal detected (Right running Fer expansion) as a function of time. The density operator,

$$\rho_R(t) = U_R(t)\rho_R(0)U_R^+(t), \tag{26}$$

and the signal detected as a function of time is formally given by,

$$Signal(t) = \langle S(t) \rangle_R = Trace\{\rho_R(t)\hat{D}\} \tag{27}$$

(b) Left running Fer expansion

$$U_L(t) = U_n(t) exp(\lambda^n F_n(t)) exp(\lambda^{n-1} F_{n-1}(t)) \ldots exp(\lambda^3 F_3(t)) exp(\lambda^2 F_2(t)) exp(\lambda F_1(t)) \tag{28}$$

Similarly, we can evaluate the density operator (Left running Fer expansion) at time t as well as derive the signal detected (Left running Fer expansion) as a function of time. The density operator,

$$\rho_L(t) = U_L(t)\rho_L(0)U_L^+(t), \tag{29}$$

and the signal detected as a function of time is also formally given by

$$Signal(t) = \langle S(t) \rangle_L = Trace\{\rho_L(t)\hat{D}\} \tag{30}$$

The $F_n(t)$ operators in the above Eqs. (25) and (28) act sequentially and are derived using the following simple integrals,

$$F_1(t) = (-i) \int_0^t H(t')dt' \tag{31-a}$$
$$F_2(t) = F_{2,0}(t) + F_{2,1}(t) + F_{2,2}(t) + \cdots \tag{31-b}$$

$$F_{2,0}(t) = \left(\frac{i}{2}\right) \int_0^t [F_1(t'), H(t')] dt' \tag{31-c}$$

$$F_{2,1}(t) = -\left(\frac{i}{3}\right) \int_0^t [F_1(t'), [F_1(t'), H(t')]] dt' \tag{31-d}$$

………………

The above Eqs. (31-a)–(31-d) show that $F_n(t)$ operators are infinite series containing progressively higher order correction terms. However, for the purpose of this article, we limit ourselves only to the lowest order correction in the operator, $F_2(t) = F_{2,0}(t)$. Although, the splitting of the time-propagator into an infinite product of exponential operators seem beneficial, from an operational aspect, the ordering of operators in the time-propagator play an important role. In the time-propagators based on Left running expansion, the $F_n(t)$ operators (of higher order 'n') act initially on the initial density operator, while in the Right running expansion, the $F_n(t)$ operators (of lower order 'n') act initially on the initial density operator. Therefore, the signal for the Right running Fer expansion can be calculated explicitly as following

$$Signal(t) = \langle S(t) \rangle_R = Trace\{\rho_R(t)\widehat{D}\} \tag{32-a}$$

$$= Tr\left[\left\{e^{F_1^R(t)} e^{F_2^R(t)} e^{F_3^R(t)} \rho(0) e^{-F_3^R(t)} e^{-F_2^R(t)} e^{-F_1^R(t)}\right\} \widehat{D}\right]$$

$$= Tr\left[\underbrace{e^{F_2^R(t)} \rho(0) e^{-F_2^R(t)}}_{\rho_{(2)}^R(t)} \underbrace{e^{F_3^R(t)} \rho(0) e^{-F_3^R(t)}}_{\rho_{(3)}^R(t)} \overbrace{e^{-F_1^R} \widehat{D} e^{F_1^R}}^{\widehat{D}_{(1)}^R(t)}\right]$$

$$= Tr\left[\rho_{(2)}^R(t) \rho_{(3)}^R(t) \widehat{D}_{(1)}^R(t)\right] \tag{32-b}$$

The signal for the Left running Fer expansion can also be calculated explicitly as

$$Signal(t) = \langle S(t) \rangle_L = Trace\{\rho_L(t)\widehat{D}\} \tag{33-a}$$

$$= Tr\left[\left\{e^{F_3^L(t)} e^{F_2^L(t)} e^{F_1^L(t)} \rho(0) e^{-F_1^L(t)} e^{-F_2^L(t)} e^{-F_3^L(t)}\right\} \widehat{D}\right]$$

$$= Tr\left[\underbrace{e^{F_2^L(t)} \rho(0) e^{-F_2^L(t)}}_{\rho_{(2)}^L(t)} \underbrace{e^{F_1^L(t)} \rho(0) e^{-F_1^L(t)}}_{\rho_{(1)}^L(t)} \overbrace{e^{-F_3^L} \widehat{D} e^{F_3^L}}^{\widehat{D}_{(3)}^L(t)}\right]$$

$$= Tr\left[\rho_{(2)}^L(t) \rho_{(1)}^L(t) \widehat{D}_{(3)}^L(t)\right] \tag{33-b}$$

The subscripts (1, 2, and 3) in the density operator $\rho(t)$ and detection operator $\widehat{D}$ represents the evolution under the corresponding $F_n^R(t)$ and $F_n^L(t)$ operators, where $F_n^R(t)$ represents the Right running counterparts of the Left running operators $F_n^L(t)$. The operators are related among them as following:

$$F_1^R(t) = F_1^L(t) = F_1(t) \tag{34-a}$$

$$F_2^R(t) = F_2^L(t) = F_2(t) \tag{34-b}$$

$$F_3^R(t) = F_3^L(t) = F_3(t) \tag{34-c}$$

As described above, the final form of the density operator at time 't' is dependent on the type of the expansion scheme (Right running or Left running) used to obtain the time-propagator. Depending on the form of the initial density operator and the detection operator, the final form of the signal expression derived from the two formulations could differ as showed in the Eqs. (32-b) and (33-b). Because most works in the literature mainly are based on Right running expansion[36,77-80], in this article, we used the Right running notation for the sake of continuity and simplicity.

## V. Application of FME to the Chemical Shielding Hamiltonian during the TOFU Pulse Sequence Radiation

The relevant spin Hamiltonian of the chemical shielding transformed into the recoupling frame takes the form[18],

$$\widetilde{H}_\sigma(t) = \omega_I(t)\left(\cos^2(Ct)I_Z + \sin(Ct)I_Y - \frac{1}{2}\sin(2Ct)I_X\right)$$
$$+ \omega_S(t)(\cos^2(Ct)S_Z + \sin(Ct)S_Y - \frac{1}{2}\sin(2Ct)S_X) \tag{35}$$

where C is an angular frequency, which is the strength of a field $B_i$ (with i = x, y, z). The functions $\omega_I(t)$ and $\omega_S(t)$ represent the time-varying chemical shifts for the two spins I and S, respectively. These functions are interaction strengths that may be expressed in terms of a Fourier series such as expressed in section II. The Interference between time dependence, $\exp(-im\omega_r t)$, of the interaction frequencies (Eq. (35)) and the time modulation given in the appendix (Eqs. (A.1) – (A.3)) arising from the TOFU pulse sequence, allows for selective recoupling of specific terms in the Hamiltonian by fulfilling the conditions

$$m\omega_r + nC = 0 \tag{36}$$

where, $m = \pm 1, \pm 2$ according to Eq. (2), and $n = \pm 1, \pm 2$. The chemical shielding Hamiltonian can be written as

$$\widetilde{H}_\sigma(t) = \sum_{m=-2}^{2} \omega_I^{(m)} \{a_X I_X + a_Y I_Y + a_Z I_Z\} + \Sigma S^{(0)} \tag{37}$$

where $\Sigma S^{(0)}$ are the sum of the equivalent terms of spin I for spin S, and the functions $a_X(t)$, $a_Y(t)$, and $a_Z(t)$ are given by

$$a_X = -\frac{1}{4i}(e^{i(m\omega_r + 2C)t} - e^{i(m\omega_r - 2C)t}) \tag{38}$$

$$a_Y = \frac{1}{2i}(e^{i(m\omega_r + C)t} - e^{i(m\omega_r - C)t}) \tag{39}$$

$$a_Z = \frac{1}{2}e^{im\omega_r t} + \frac{1}{4}(e^{i(m\omega_r + 2C)t} + e^{i(m\omega_r - 2C)t}) \tag{40}$$

The dipolar coupling Hamiltonian takes the form[18],

$$\tilde{H}_{IS}(t) = \frac{3}{2}\omega_{IS}(t)\begin{Bmatrix} \cos^4(Ct)2I_ZS_Z + \sin^2(Ct)\cos^2(Ct)2I_XS_X \\ +\sin^2(Ct)2I_YS_Y + \frac{1}{2}\sin(2Ct)\left[\cos(Ct)(2I_ZS_Y + 2I_YS_Z)\right. \\ \left. -\sin(Ct)(2I_XS_Y + 2I_YS_X) - \cos^2(Ct)(2I_ZS_X + 2I_XS_Z\right] \end{Bmatrix} \quad (41)$$

where components from the invariant (and non-recoupled) $\vec{I}.\vec{S}$ part of the Hamiltonian are ignored. In this article, we only considered and treated the chemical shielding Hamiltonian and we leaved the dipolar coupling Hamiltonian, which will be treated in a forthcoming publication.

The first order term of the FME can be calculated as

$$F_1^\sigma = \frac{1}{T}\int_0^T \tilde{H}(\tau)d\tau = \frac{1}{T}\int_0^T \Big\{ \overbrace{\omega_I(t)\cos^2(Ct)\,I_Z}^{I_1} + \underbrace{\omega_I(t)\sin(Ct)\,I_Y}_{I_2} -$$

$$\overbrace{\frac{1}{2}\omega_I(t)\sin(2Ct)\,I_X}^{I_3} + \overbrace{\omega_S(t)\cos^2(Ct)\,S_Z}^{I_4} + \underbrace{\omega_S(t)\sin(Ct)\,S_Y}_{I_5} - \overbrace{\frac{1}{2}\omega_S(t)\sin(2Ct)\,S_X}^{I_6} \Big\}dt$$

(42)

where the integrals $I_1$, $I_2$, … $I_6$ are calculated in the appendix. In respect with the Dirac function integration[23],

$$\frac{1}{T}\int_0^T e^{im\omega_r t}dt = \delta_{m,0} \quad (43)$$

we have

$$I_1 = \sum_{m=-2}^{2} \omega_I^{(m)} \begin{cases} \frac{1}{2}, & \text{for } m\omega_r = 0 \\ \frac{1}{4}, & \text{for } m\omega_r + 2C = 0 \\ \frac{1}{4}, & \text{for } m\omega_r - 2C = 0 \end{cases} \quad (44\text{-}1)$$

$$I_2 = \sum_{m=-2}^{2} \omega_I^{(m)} \begin{cases} \frac{1}{2i}, & \text{for } m\omega_r + C = 0 \\ \frac{1}{-2i}, & \text{for } m\omega_r - C = 0 \end{cases} \quad (44\text{-}2)$$

$$I_3 = \sum_{m=-2}^{2} \omega_I^{(m)} \begin{cases} \frac{-1}{4i}, & \text{for } 2C + m\omega_r = 0 \\ \frac{1}{4i}, & \text{for } 2C - m\omega_r = 0 \end{cases} \quad (44\text{-}3)$$

$$I_4 = \Sigma_{m=-2}^{2} \omega_S^{(m)} \begin{cases} \frac{1}{2}, & for\ m\omega_r = 0 \\ \frac{1}{4}, & for\ m\omega_r + 2C = 0 \\ \frac{1}{4}, & for\ m\omega_r - 2C = 0 \end{cases} \qquad (44\text{-}4)$$

$$I_5 = \Sigma_{m=-2}^{2} \omega_S^{(m)} \begin{cases} \frac{1}{2i}, & for\ m\omega_r + C = 0 \\ \frac{1}{-2i}, & for\ m\omega_r - C = 0 \end{cases} \qquad (44\text{-}5)$$

$$I_6 = \Sigma_{m=-2}^{2} \omega_S^{(m)} \begin{cases} \frac{-1}{4i}, & for\ 2C + m\omega_r = 0 \\ \frac{1}{4i}, & for\ 2C - m\omega_r = 0 \end{cases} \qquad (44\text{-}6)$$

We consider selective recoupling of specific terms in the toggling Hamiltonian (Eq. (35)) by fulfilling the conditions of the Eq. (36). We considered two classes of selective recoupling experiments obtained using,

(a) $C = \frac{1}{2}m\omega_r$, $\qquad (45)$

and

(b) $C = m\omega_r$. $\qquad (46)$

For $\boldsymbol{C = \frac{1}{2}m\omega_r}$, we have the first order term of the FME calculated as,

$$F_1^\sigma = \left(\frac{1}{4}I_Z + \frac{1}{4i}I_X\right) \Sigma_{m=-2}^{2} \omega_I^{(m)} + \left(\frac{1}{4}S_Z + \frac{1}{4i}S_X\right) \Sigma_{m=-2}^{2} \omega_S^{(m)} \qquad (47)$$

and for $\boldsymbol{C = m\omega_r}$, we have the first order term of the FME calculated as,

$$F_1^\sigma = -\frac{1}{2i}I_Y \Sigma_{m=-2}^{2} \omega_I^{(m)} - \frac{1}{2i}S_Y \Sigma_{m=-2}^{2} \omega_S^{(m)} \qquad (48)$$

The associate transformation results from

$$\Lambda_1(t) = \int_0^t \widetilde{H}(\tau)d\tau - tF_1 \qquad (49)$$

where we choose $\Lambda_1(0) = 0$. After calculations showed in the appendix, we obtained

$$\Lambda_1(t)$$

$$= \sum_{m=-2}^{2} \omega_I^{(m)} \begin{cases} [\frac{1}{2im\omega_r}(e^{im\omega_r t} - 1) + \frac{1}{4i(m\omega_r + 2C)}(e^{i(m\omega_r+2C)t} - 1) \\ + \frac{1}{4i(m\omega_r - 2C)}(e^{i(m\omega_r-2C)t} - 1)]I_Z + [\frac{-1}{2(m\omega_r + C)}(e^{i(m\omega_r+C)t} - 1) \\ + \frac{1}{2(m\omega_r - C)}(e^{i(m\omega_r-C)t} - 1)]I_Y + [\frac{1}{4(2C + m\omega_r)}(e^{i(2C+m\omega_r)t} - 1) \\ - \frac{1}{4(m\omega_r - 2C)}(e^{i(m\omega_r-2C)t} - 1)]I_X \end{cases}$$

$$+ \Sigma S \text{ - } tF_1 \tag{50}$$

where $\Sigma S$ are the sum of the equivalent terms of spin I for spin S. If we choose $\boldsymbol{C = \frac{1}{2}m\omega_r}$, corresponding also to $\boldsymbol{m\omega_r - 2C = 0}$, the first-order term of the argument of the operator evolution can be evaluated as

$$\Lambda_1(t) = \sum_{m=-2}^{2} \omega_I^{(m)}\{a_{IX}(t)I_X + a_{IY}(t)I_Y + a_{IZ}(t)I_Z\} + \Sigma S^{(1)}, \tag{51}$$

where $\Sigma S^{(1)}$ are the sum of the equivalent terms of spin I for spin S, and the functions $a_{IX}(t)$, $a_{IY}(t)$, and $a_{IZ}(t)$ are given by

$$a_{IX}(t) = \frac{1}{(8m\omega_r)}(e^{i(2m\omega_r)t} - 1), \tag{52}$$

$$a_{IY}(t) = \frac{-1}{(3m\omega_r)}\left(e^{i(\frac{3}{2}m\omega_r)t} - 1\right) + \frac{1}{(m\omega_r)}(e^{i(\frac{1}{2}m\omega_r)t} - 1) \tag{53}$$

$$a_{IZ}(t) = \frac{1}{2im\omega_r}(e^{im\omega_r t} - 1) + \frac{1}{8i(m\omega_r)}(e^{i(2m\omega_r)t} - 1) \tag{54}$$

If we choose $\boldsymbol{C = m\omega_r}$, corresponding to $\boldsymbol{m\omega_r - C = 0}$, the first-order term of the argument of the operator evolution can be evaluated as

$$\Lambda_{1,1}(t) = \sum_{m=-2}^{2} \omega_I^{(m)}\{a_{IX,1}(t)I_X + a_{IY,1}(t)I_Y + a_{IZ,1}(t)I_Z\} + \Sigma S^{(2)}, \tag{55}$$

where $\Sigma S^{(2)}$ are the sum of the equivalent terms of spin I for spin S, and the functions $a_{IX,1}(t)$, $a_{IY,1}(t)$, and $a_{IZ,1}(t)$ are given by

$$a_{IX,1}(t) = \frac{1}{12m\omega_r}\left(e^{i(3m\omega_r)t} - 1\right) + \frac{1}{4m\omega_r}(e^{-im\omega_r t} - 1), \tag{56}$$

$$a_{IY,1}(t) = \frac{-1}{4m\omega_r}\left(e^{i(2m\omega_r)t} - 1\right), \tag{57}$$

$$a_{IZ,1}(t) = \frac{1}{2im\omega_r}\left(e^{im\omega_r t} - 1\right) + \frac{1}{12im\omega_r}\left(e^{i(3m\omega_r)t} - 1\right) - \frac{1}{4im\omega_r}\left(e^{-im\omega_r t} - 1\right). \tag{58}$$

High order terms can also be obtained after lengthy calculations and the details of calculations are shown in the appendix. The second order terms are computed to be for $C = \frac{1}{2}m\omega_r$,

$$F_2^\sigma = \frac{1}{2iT}\int_0^T[\widetilde{H}(\tau) + F_1, \Lambda_1(\tau)]d\tau = (I_1' + I_2' + I_5')I_X + (I_3' + I_4' + I_7' + I_8')I_Y \quad (59)$$
$$+ (I_6' + I_9' + I_{10}')I_Z + \Sigma S^{(3)},$$

where $\Sigma S^{(3)}$ are the sum of the equivalent terms of spin I for spin S, and the integrals, $I_1'$, $I_2', I_3', \ldots I_9'$, and $I_{10}'$ are given in the following and are calculated in the appendix,

$$I_1' = \frac{-1}{2T}\int_0^T a_{IY}\omega_I(t)\cos^2(Ct)\,dt$$

$$= -\frac{1}{2}\sum_{m=-2}^{2}\omega_I^{(m)}\begin{cases} 0 & \text{for } m\omega_r = 0 \\ -\frac{1}{12m\omega_r} & \text{for } \frac{5}{2}m\omega_r + 2C = 0 \\ -\frac{1}{6m\omega_r} & \text{for } m\omega_r + 2C = 0 \\ \frac{1}{4m\omega_r} & \text{for } \frac{3}{2}m\omega_r + 2C = 0 \\ -\frac{1}{12m\omega_r} & \text{for } \frac{5}{2}m\omega_r - 2C = 0 \\ -\frac{1}{6m\omega_r} & \text{for } m\omega_r - 2C = 0 \\ \frac{1}{4m\omega_r} & \text{for } \frac{3}{2}m\omega_r - 2C = 0 \end{cases} \quad (60)$$

$$I_2' = \frac{-1}{8T}\sum_{m=-2}^{2}\omega_I^{(m)}\int_0^T a_{IY}\,dt$$
$$= -\frac{1}{8}\sum_{m=-2}^{2}\omega_I^{(m)}\begin{cases} 0 & \text{for } m\omega_r = 0 \\ \frac{-2}{3m\omega_r} & \text{for } m\omega_r \neq 0 \end{cases} \quad (61)$$

$$I_3' = \frac{1}{2T}\int_0^T a_{IX}\omega_I(t)\cos^2(Ct)\,dt$$

$$= \frac{1}{16m\omega_r}\sum_{m=-2}^{2}\omega_I^{(m)}\begin{cases} 0 & \text{for } m\omega_r = 0 \\ \frac{1}{4} & \text{for } 3m\omega_r + 2C = 0 \\ -\frac{1}{4} & \text{for } m\omega_r + 2C = 0 \\ \frac{1}{4} & \text{for } 3m\omega_r - 2C = 0 \\ -\frac{1}{4} & \text{for } m\omega_r - 2C = 0 \end{cases} \quad (62)$$

$$I_4' = \frac{1}{8T}\sum_{m=-2}^{2}\omega_I^{(m)}\int_0^T a_{IX}\,dt$$
$$= \frac{1}{8}\sum_{m=-2}^{2}\omega_I^{(m)}\begin{cases} 0 & \text{for } m\omega_r = 0 \\ \frac{-1}{8m\omega_r} & \text{for } m\omega_r \neq 0 \end{cases} \quad (63)$$

$$I_5' = \frac{1}{2T}\int_0^T a_{IZ}\omega_I(t)\sin(Ct)\,dt$$

$$= \frac{1}{4i}\sum_{m=-2}^{2}\omega_I^{(m)}\begin{cases} \frac{1}{2im\omega_r} & \text{for } 2m\omega_r + C = 0 \\ \frac{-5}{8im\omega_r} & \text{for } m\omega_r + C = 0 \\ -\frac{1}{2im\omega_r} & \text{for } 2m\omega_r - C = 0 \\ \frac{5}{8im\omega_r} & \text{for } m\omega_r - C = 0 \\ -\frac{1}{8im\omega_r} & \text{for } 3m\omega_r - C = 0 \\ \frac{1}{8im\omega_r} & \text{for } 3m\omega_r + C = 0 \end{cases} \quad (64)$$

$$I_6' = \frac{-1}{2T}\int_0^T a_{IX}\omega_I(t)\sin(Ct)\,dt$$

$$= \frac{-1}{32im\omega_r}\sum_{m=-2}^{2}\omega_I^{(m)}\begin{cases} 1 & \text{for } 3m\omega_r + C = 0 \\ -1 & \text{for } m\omega_r + C = 0 \\ -1 & \text{for } 3m\omega_r - C = 0 \\ 1 & \text{for } m\omega_r - C = 0 \end{cases} \quad (65)$$

$$I_7' = \frac{1}{4T}\int_0^T a_{IZ}\omega_I(t)\sin(2Ct)\,dt$$

$$= \frac{1}{8i}\sum_{m=-2}^{2}\omega_I^{(m)}\begin{cases} \frac{1}{2im\omega_r} & \text{for } 2m\omega_r + 2C = 0 \\ -\frac{5}{8im\omega_r} & \text{for } m\omega_r + 2C = 0 \\ \frac{1}{8im\omega_r} & \text{for } 3m\omega_r + 2C = 0 \\ \frac{-1}{2im\omega_r} & \text{for } 2m\omega_r - 2C = 0 \\ \frac{1}{2im\omega_r} & \text{for } m\omega_r - 2C = 0 \\ -\frac{1}{8im\omega_r} & \text{for } 3m\omega_r - 2C = 0 \\ \frac{1}{8im\omega_r} & \text{for } m\omega_r - 2C = 0 \end{cases} \quad (66)$$

$$I_8' = \frac{-1}{8iT}\sum_{m=-2}^{2}\omega_I^{(m)}\int_0^T a_{IZ}\,dt$$

$$= \frac{1}{8i}\sum_{m=-2}^{2}\omega_I^{(m)}\begin{cases} 0 & \text{for } m\omega_r = 0 \\ \frac{-5}{8im\omega_r} & \text{for } m\omega_r \neq 0 \end{cases} \quad (67)$$

$$I_9' = \frac{-1}{4T}\int_0^T a_{IY}\omega_I(t)\sin(2Ct)\,dt$$

$$= \frac{1}{4i}\sum_{m=-2}^{2}\omega_I^{(m)}\begin{cases}\frac{-1}{3m\omega_r} & for\ \frac{5}{2}m\omega_r + 2C = 0\\ \frac{-2}{3m\omega_r} & for\ m\omega_r + 2C = 0\\ \frac{1}{m\omega_r} & for\ \frac{3}{2}m\omega_r + 2C = 0\\ \frac{1}{3m\omega_r} & for\ \frac{5}{2}m\omega_r - 2C = 0\\ \frac{2}{3m\omega_r} & for\ m\omega_r - 2C = 0\\ \frac{-1}{m\omega_r} & for\ \frac{3}{2}m\omega_r - 2C = 0\end{cases} \tag{68}$$

$$I'_{10} = \frac{1}{8iT}\sum_{m=-2}^{2}\omega_I^{(m)}\int_0^T a_{IY}\,dt$$

$$= \frac{1}{8i}\sum_{m=-2}^{2}\omega_I^{(m)}\begin{cases}0 & for\ m\omega_r = 0\\ \frac{-2}{3m\omega_r} & for\ m\omega_r \neq 0\end{cases} \tag{69}$$

For $C = \frac{1}{2}m\omega_r$, we have the second order term of the FME calculated as,

$$F_2^\sigma = \sum_{m=-2}^{2}\omega_I^{(m)}\left(\frac{1}{2m\omega_r}\right)\left[\frac{1}{3}I_X - \frac{1}{4}I_Y - \frac{1}{3i}I_Z\right] + \sum S^{(3)} \tag{70-1}$$

The propagator derived from the FME can be written as

$$U(\tau_C) \approx exp\{-i\tau_C(F_1^\sigma + F_2^\sigma + \cdots)\} \tag{70-2}$$

where $F_1^\sigma, F_2^\sigma, \ldots etc$ correspond to the 1st, 2nd … etc… Floquet operator orders, respectively.

For $C = m\omega_r$,

$$F_{2,1}^\sigma = \frac{1}{2iT}\int_0^T[\widetilde{H}(\tau) + F_1, \Lambda_{1,1}(\tau)]d\tau = (I'_{1,1} + I'_{2,1} + I'_{5,1})I_X + (I'_{3,1} + I'_{4,1} + I'_{7,1} + I'_{8,1})I_Y + (I'_{6,1} + I'_{9,1} + I'_{10,1})I_Z + \sum S^{(4)}, \tag{71}$$

where $\Sigma S^{(4)}$ are the sum of the equivalent terms of spin I for spin S, and the integrals, $I'_{1,1}$, $I'_2, I'_{3,1} \ldots I'_{9,1}$, and $I'_{10,1}$ are calculated in the following,

$$I'_{1,1} = \frac{-1}{2T}\int_0^T a_{IY,1}\omega_I(t)cos^2(Ct)\,dt \tag{72}$$

$$= \frac{1}{8m\omega_r}\Sigma_{m=-2}^{2}\omega_I^{(m)}\begin{cases} 0 & for\ m\omega_r = 0 \\ \frac{1}{4} & for\ 3m\omega_r + 2C = 0 \\ \frac{1}{4} & for\ 3m\omega_r - 2C = 0 \\ \frac{-1}{4} & for\ m\omega_r + 2C = 0 \\ -\frac{1}{4} & for\ m\omega_r - 2C = 0 \end{cases}$$

$$I'_{2,1} = \frac{-1}{8T}\Sigma_{m=-2}^{2}\omega_I^{(m)}\int_0^T a_{IY,1}\,dt$$
$$= \Sigma_{m=-2}^{2}\omega_I^{(m)}\begin{cases} 0 & for\ m\omega_r = 0 \\ \frac{-1}{32m\omega_r} & for\ m\omega_r \neq 0 \end{cases} \quad (73)$$

$$I'_{3,1} = \frac{1}{2T}\int_0^T a_{IX,1}\omega_I(t)\cos^2(Ct)\,dt$$
$$=$$
$$\frac{1}{8m\omega_r}\Sigma_{m=-2}^{2}\omega_I^{(m)}\begin{cases} 0 & for\ m\omega_r = 0 \\ \frac{7}{12} & for\ 4m\omega_r + 2C = 0 \\ \frac{1}{6} & for\ m\omega_r + 2C = 0 \\ \frac{7}{2} & for\ 4m\omega_r - 2C = 0 \\ \frac{1}{6} & for\ m\omega_r - 2C = 0 \end{cases} \quad (74)$$

$$I'_{4,1} = \frac{1}{8T}\Sigma_{m=-2}^{2}\omega_I^{(m)}\int_0^T a_{IX,1}\,dt$$
$$= \frac{1}{32}\Sigma_{m=-2}^{2}\omega_I^{(m)}\begin{cases} 0 & for\ m\omega_r = 0 \\ \frac{-5}{4m\omega_r} & for\ m\omega_r \neq 0 \end{cases} \quad (75)$$

$$I'_{5,1} = \frac{1}{2T}\int_0^T a_{IZ,1}\omega_I(t)\sin(Ct)\,dt$$
$$= \frac{-1}{8m\omega_r}\Sigma_{m=-2}^{2}\omega_I^{(m)}\begin{cases} 1 & for\ 2m\omega_r + C = 0 \\ \frac{-2}{3} & for\ m\omega_r + C = 0 \\ \frac{1}{6} & for\ 4m\omega_r + C = 0 \\ -1 & for\ 2m\omega_r - C = 0 \\ \frac{2}{3} & for\ m\omega_r - C = 0 \\ \frac{-1}{6} & for\ 4m\omega_r - C = 0 \end{cases} \quad (76)$$

$$I'_6 = \frac{-1}{2T}\int_0^T a_{IX,1}\omega_I(t)\sin(Ct)\,dt$$

$$= \frac{-1}{16im\omega_r}\sum_{m=-2}^{2}\omega_I^{(m)}\begin{cases}\frac{1}{3} & \text{for } 4m\omega_r + C = 0 \\ -\frac{4}{3} & \text{for } m\omega_r + C = 0 \\ -\frac{1}{3} & \text{for } 4m\omega_r - C = 0 \\ \frac{4}{3} & \text{for } m\omega_r - C = 0\end{cases} \quad (77)$$

$$I'_{7,1} = \frac{1}{4T}\int_0^T a_{IZ,1}\omega_I(t)\sin(2Ct)\,dt$$

$$= \frac{1}{16im\omega_r}\sum_{m=-2}^{2}\omega_I^{(m)}\begin{cases}1 & \text{for } 2m\omega_r + 2C = 0 \\ \frac{-2}{3} & \text{for } m\omega_r + 2C = 0 \\ \frac{1}{6} & \text{for } 4m\omega_r + 2C = 0 \\ -1 & \text{for } 2m\omega_r - 2C = 0 \\ \frac{2}{3} & \text{for } m\omega_r - 2C = 0 \\ \frac{-1}{6} & \text{for } 4m\omega_r - 2C = 0\end{cases} \quad (78)$$

$$I'_{8,1} = \frac{-1}{8iT}\sum_{m=-2}^{2}\omega_I^{(m)}\int_0^T a_{IZ,1}\,dt$$

$$= \frac{1}{16m\omega_r}\sum_{m=-2}^{2}\omega_I^{(m)}\begin{cases}0 & \text{for } m\omega_r = 0 \\ \frac{-2}{3} & \text{for } m\omega_r \neq 0\end{cases} \quad (79)$$

$$I'_{9,1} = \frac{-1}{4T}\int_0^T a_{IY,1}\omega_I(t)\sin(2Ct)\,dt$$

$$= \frac{1}{32im\omega_r}\sum_{m=-2}^{2}\omega_I^{(m)}\begin{cases}1 & \text{for } 3m\omega_r + 2C = 0 \\ -1 & \text{for } 3m\omega_r - 2C = 0 \\ -1 & \text{for } m\omega_r + 2C = 0 \\ 1 & \text{for } m\omega_r - 2C = 0\end{cases} \quad (80)$$

$$I'_{10,1} = \frac{1}{8iT}\sum_{m=-2}^{2}\omega_I^{(m)}\int_0^T a_{IY,1}\,dt$$

$$= \frac{-1}{32im\omega_r}\sum_{m=-2}^{2}\omega_I^{(m)}\begin{cases}0 & \text{for } m\omega_r = 0 \\ -1 & \text{for } m\omega_r \neq 0\end{cases} \quad (81)$$

For $C = m\omega_r$, we have the second order term of the FME calculated as,

$$F_{2,1}^\sigma = \sum_{m=-2}^{2}\omega_I^{(m)}\left(\frac{1}{m\omega_r}\right)\left[-\frac{11}{96}I_X - \left(\frac{31}{384} + \frac{1}{16i}\right)I_Y - \frac{5}{96i}I_Z\right] + \sum S^{(4)} \quad (82)$$

The propagator derived from the FME can be written as

$$U(\tau_C) \approx \exp\{-i\tau_C(F_1^\sigma + F_{2,1}^\sigma + \cdots)\} \quad (83)$$

where $F_1^\sigma, F_{2,1}^\sigma, \dots etc$ correspond to the 1st, 2nd ... etc... Floquet operator orders, respectively.

## VI. Application of FE to the Chemical Shielding Hamiltonian during the TOFU Pulse Sequence Radiation

Here, we examine the Fer expansion by using the TOFU pulse sequence radiation. Using the chemical shielding Hamiltonian when transformed into the recoupling frame defined by the Eq. (35), we first calculate the function

$$F_1(t) = -i \int_0^t \widetilde{H}_\sigma(\tau) d\tau = \sum_{m=-2}^{2} \omega_I^{(m)}\{(f_x(t))I_X + (f_y(t))I_Y + (f_z(t))I_Z\} + \Sigma S^{(5)} \quad (84)$$

where $\Sigma S^{(5)}$ are the sum of the equivalent terms of spin I for spin S, and the functions, $f_x(t), f_y(t)$, and $f_z(t)$, are given by,

$$f_x(t) = \frac{-i}{4(2C+m\omega_r)}\left(e^{i(2C+m\omega_r)t} - 1\right) + \frac{i}{4(m\omega_r-2C)}(e^{i(m\omega_r-2C)t} - 1), \quad (85)$$

$$f_y(t) = \frac{i}{2(m\omega_r+C)}\left(e^{i(m\omega_r+C)t} - 1\right) - \frac{i}{2(m\omega_r-C)}(e^{i(m\omega_r-C)t} - 1) \quad (86)$$

$$f_z(t) = \frac{-1}{2m\omega_r}\left(e^{im\omega_r t} - 1\right) - \frac{1}{4(m\omega_r+2C)}\left(e^{i(m\omega_r+2C)t} - 1\right) - \frac{1}{4(m\omega_r-2C)}(e^{i(m\omega_r-2C)t} - 1) \quad (87)$$

If we choose $C = \frac{1}{2}m\omega_r$, we have the function,

$$F_1(t) = \sum_{m=-2}^{2} \omega_I^{(m)}\{(f_{1x}(t))I_X + (f_{1y}(t))I_Y + (f_{1z}(t))I_Z\} + \Sigma S^{(6)} \quad (88)$$

where $\Sigma S^{(6)}$ are the sum of the equivalent terms of spin I for spin S, and the functions, $f_{1x}(t), f_{1y}(t)$, and $f_{1z}(t)$, are given by,

$$f_{1x}(t) = \frac{-i}{(8m\omega_r)}\left(e^{i(2m\omega_r)t} - 1\right) - \frac{1}{4}t \quad (89)$$

$$f_{1y}(t) = \frac{i}{3m\omega_r}\left(e^{i\left(\frac{3}{2}m\omega_r\right)t} - 1\right) - \frac{i}{m\omega_r}(e^{i\left(\frac{1}{2}m\omega_r\right)t} - 1) \quad (90)$$

$$f_{1z}(t) = \frac{-1}{2m\omega_r}\left(e^{im\omega_r t} - 1\right) - \frac{1}{8m\omega_r}\left(e^{i(2m\omega_r)t} - 1\right) - \frac{i}{4}t \quad (91)$$

Using the condition $T = \tau_C = \frac{2\pi}{\omega_r}$, we have

$$F_1(\tau_C) = \sum_{m=-2}^{2} \omega_I^{(m)}\left\{(-\frac{1}{4}\tau_C)I_X + (-\frac{i}{4}\tau_C)I_Z\right\} + \Sigma S^{(6)} \quad (92)$$

which leads to the zero-order average Hamiltonian for the Fer expansion

$$\bar{H}_{Fer}^{(0)} = \sum_{m=-2}^{2} \omega_I^{(m)} \left\{ \left(-\frac{i}{4}\right)I_X + \left(\frac{1}{4}\right)I_Z \right\} + \sum_{m=-2}^{2} \omega_S^{(m)} \left\{ \left(-\frac{i}{4}\right)S_X + \left(\frac{1}{4}\right)S_Z \right\} \tag{93}$$

If we choose $\mathbf{C} = \mathbf{m\omega_r}$, we have the function

$$F_1(t) = \sum_{m=-2}^{2} \omega_I^{(m)} \{(f_{2x}(t))I_X + (f_{2y}(t))I_Y + (f_{2z}(t))I_Z\} + \Sigma S^{(7)} \tag{94}$$

where $\Sigma S^{(7)}$ are the sum of the equivalent terms of spin I for spin S, and the functions, $f_{2x}(t)$, $f_{2y}(t)$, and $f_{2z}(t)$, are given by,

$$f_{2x}(t) = \frac{-i}{12m\omega_r}\left(e^{i(3m\omega_r)t} - 1\right) - \frac{1}{4m\omega_r}(e^{-im\omega_r t} - 1) \tag{95}$$

$$f_{2y}(t) = \frac{i}{4m\omega_r}\left(e^{i(2m\omega_r)t} - 1\right) + \frac{1}{2}t \tag{96}$$

$$f_{2z}(t) = \frac{-1}{2m\omega_r}\left(e^{im\omega_r t} - 1\right) - \frac{1}{12m\omega_r}\left(e^{i(3m\omega_r)t} - 1\right) + \frac{1}{4m\omega_r}(e^{-im\omega_r t} - 1) \tag{97}$$

We have,

$$F_1(\tau_C) = \sum_{m=-2}^{2} \omega_I^{(m)} \left\{\frac{1}{2}\tau_C\right\} I_Y + \sum_{m=-2}^{2} \omega_S^{(m)} \left\{\frac{1}{2}\tau_C\right\} S_Y \tag{98}$$

which lead to the zero-order average Hamiltonian for the Fer expansion

$$\bar{H}_{Fer}^{(0)} = \sum_{m=-2}^{2} \omega_I^{(m)} \left\{\left(\frac{i}{2}\right)I_Y\right\} + \sum_{m=-2}^{2} \omega_S^{(m)} \left\{\left(\frac{i}{2}\right)S_Y\right\} \tag{99}$$

Using the time modulation given in the Appendix, Eq. (A.1), the Hamiltonian takes the form

$$\tilde{H}_\sigma(t) = a_X(t)I_X + a_Y(t)I_Y + a_Z(t)I_Z + \Sigma S^{(8)} \tag{100}$$

where $\Sigma S^{(8)}$ are the sum of the equivalent terms of spin I for spin S, and the functions, $a_X(t)$, $a_Y(t)$, and $a_Z(t)$, are given by,

$$a_X(t) = -\frac{1}{2}\sum_{m=-2}^{2} \omega_I^{(m)} \frac{1}{2i}(e^{i(2C+m\omega_r)t} - e^{i(m\omega_r - 2C)t}) \tag{101}$$

$$a_Y(t) = \sum_{m=-2}^{2} \omega_I^{(m)} \frac{1}{2i}(e^{i(C+m\omega_r)t} - e^{i(m\omega_r - C)t}) \tag{102}$$

$$a_Z(t) = \sum_{m=-2}^{2} \omega_I^{(m)} \left(\frac{1}{2}e^{i(m\omega_r)t} + \frac{1}{4}e^{i(m\omega_r + 2C)t} + \frac{1}{4}e^{i(m\omega_r - 2C)t}\right) \tag{103}$$

The major term in the first-order term is calculated.

For $C = \frac{1}{2}m\omega_r$, we have

$$H_{1,0}(t) = -\frac{1}{2}[F_1(t), \tilde{H}_\sigma(t)] = -\frac{1}{2}i\{(f_{1y}(t)a_Z(t) - f_{1z}(t)a_Y(t))I_X + (f_{1z}(t)a_X(t) - f_{1x}(t)a_Z(t))I_Y + (f_{1x}(t)a_Y(t) - f_{1y}(t)a_X(t))I_Z\} \tag{104}$$

and the correspond term

$$F_{2,0}(t) = -i\int_0^t H_{1,0}(t')dt' = F_{2,0}^X(t) + F_{2,0}^Y(t) + F_{2,0}^Z(t) + \Sigma S^{(9)} \tag{105}$$

where $\Sigma S^{(9)}$ are the sum of the equivalent terms of spin I for spin S, and the functions, $F_{2,0}^X(t)$, $F_{2,0}^Y(t)$, and $F_{2,0}^Z(t)$, are calculated and results are given below. For simplicity reasons without losing the generality of the problem, we have set, $m = m'$, and obtained,

$$F_{2,0}^X(t) = -\frac{1}{2}(\Sigma_{m=-2}^2 \omega_I^{(m)})^2 \left\{ \begin{array}{l} \frac{1}{6m^2\omega_r^2}\left[\frac{2}{5}\left(e^{i\frac{5}{2}m\omega_r t} - 1\right) - \left(e^{im\omega_r t} - 1\right)\right] - \frac{1}{2m^2\omega_r^2} * \\ * \left[\frac{2}{3}\left(e^{i\frac{3}{2}m\omega_r t} - 1\right) - \left(e^{im\omega_r t} - 1\right)\right] - \frac{1}{4m^2\omega_r^2}\left[\frac{2}{5}\left(e^{i\frac{5}{2}m\omega_r t} - 1\right) - \frac{1}{2}\left(e^{i2m\omega_r t} - 1\right)\right] \\ -\frac{1}{4m\omega_r}\left[\frac{2}{m\omega_r}\left(e^{i\frac{1}{2}m\omega_r t} - 1\right)\right] \\ -it \\ -\frac{1}{2m^2\omega_r^2}\left[\frac{1}{5}\left(e^{i\frac{5}{2}m\omega_r t} - 1\right) - \frac{1}{3}\left(e^{i\frac{3}{2}m\omega_r t} - 1\right)\right] \\ -\frac{1}{8m^2\omega_r^2}\left[\frac{1}{7}(e^{i\frac{7}{2}m\omega_r t} - 1)\right. \\ \left. -\frac{1}{3}(e^{i\frac{3}{2}m\omega_r t} - 1)\right] + \frac{i}{9m^2\omega_r^2}\left[\left(i\frac{3}{2}m\omega_r t - 1\right)e^{i\frac{3}{2}m\omega_r t} + 1\right] \\ +\frac{1}{2m^2\omega_r^2}\left[\frac{1}{3}\left(e^{i\frac{3}{2}m\omega_r t} - 1\right) - \left(e^{i\frac{1}{2}m\omega_r t} - 1\right)\right] + \frac{1}{8m^2\omega_r^2}\left[\frac{1}{5}\left(e^{i\frac{5}{2}m\omega_r t} - 1\right) \right. \\ \left. -\left(e^{i\frac{1}{2}m\omega_r t} - 1\right)\right] \\ -\frac{i}{m^2\omega_r^2}\left[\left(i\frac{m}{2}\omega_r t - 1\right)e^{i\frac{m}{2}\omega_r t} + 1\right] \end{array} \right\} I_X \tag{106}$$

which is simplified after calculation

$$F_{2,0}^X(t) = -\frac{1}{2}\left(\Sigma_{m=-2}^2 \omega_I^{(m)}\right)^2 \left(\frac{1}{m^2\omega_r^2}\right)\left\{-\frac{1}{56}e^{i\frac{7}{2}m\omega_r t} - \frac{13}{120}e^{i\frac{5}{2}m\omega_r t} + \left(\frac{1}{24} + \frac{i}{9}\left(i\frac{3}{2}m\omega_r t - 1\right)\right)e^{i\frac{3}{2}m\omega_r t} - \left(\frac{3}{8} + i\left(i\frac{m}{2}\omega_r t - 1\right)\right)e^{i\frac{m}{2}\omega_r t} + \frac{1}{8}e^{i2m\omega_r t} + \frac{1}{3}e^{im\omega_r t} - \frac{8}{9}i - it + \frac{631}{840}\right\}I_X \tag{107}$$

We can calculate the expression at $\tau_C = \frac{2\pi}{\omega_r}$,

$$F_{2,0}^X(\tau_C) = -\frac{1}{2}\left(\sum_{m=-2}^{2}\omega_I^{(m)}\right)^2 \left(\frac{1}{m^2\omega_r^2}\right)\left\{-\frac{1}{56}e^{i7m\pi} - \frac{13}{120}e^{i5m\pi} + \left(\frac{1}{24} + \frac{i}{9}(i3m\pi - 1)\right)e^{i3m\pi} - \left(\frac{3}{8} + i(im\pi - 1)\right)e^{im\pi} + \frac{1}{8}e^{i4m\pi} + \frac{1}{3}e^{i2m\pi} - \frac{8}{9}i - i\frac{2\pi}{\omega_r} + \frac{631}{840}\right\}I_X \quad (108)$$

and

$$\frac{iF_{2,0}^X(\tau_C)}{\tau_C} = -\frac{1}{2}\left(\sum_{m=-2}^{2}\omega_I^{(m)}\right)^2 \left(\frac{i}{2\pi m^2\omega_r}\right)\left\{-\frac{1}{56}e^{i7m\pi} - \frac{13}{120}e^{i5m\pi} + \left(\frac{1}{24} + \frac{i}{9}(i3m\pi - 1)\right)e^{i3m\pi} - \left(\frac{3}{8} + i(im\pi - 1)\right)e^{im\pi} + \frac{1}{8}e^{i4m\pi} + \frac{1}{3}e^{i2m\pi} - \frac{8}{9}i - i\frac{2\pi}{\omega_r} + \frac{631}{840}\right\}I_X = A_X I_X \quad (109)$$

$F_{2,0}^Y(t)$

$$= \frac{1}{2}\left(\sum_{m=-2}^{2}\omega_I^{(m)}\right)^2 \left\{\begin{array}{c} \frac{1}{8im^2\omega_r^2}\left[\frac{1}{3}(e^{i3m\omega_r t} - 1) - \frac{1}{2}(e^{i2m\omega_r t} - 1)\right] + \frac{1}{32im^2\omega_r^2} * \\ *\left[\begin{array}{c}\frac{1}{4}(e^{i4m\omega_r t} - 1) \\ -\frac{1}{2}(e^{i2m\omega_r t} - 1)\end{array}\right] + \frac{17}{64im^2\omega_r^2}\left[1 + (i2m\omega_r t - 1)e^{i2m\omega_r t}\right] \\ -\frac{1}{8m\omega_r}\left[\begin{array}{c}\frac{1}{im\omega_r}(e^{im\omega_r t} - 1) \\ -t\end{array}\right] \\ -\frac{1}{32m\omega_r}\left[\frac{1}{i2m\omega_r}(e^{i2m\omega_r t} - 1) - t\right] \\ -\frac{1}{16im^2\omega_r^2}\left[\frac{1}{3}(e^{i3m\omega_r t} - 1) - (e^{im\omega_r t} - 1)\right] - \frac{1}{32im^2\omega_r^2}\left[\frac{1}{4}(e^{i4m\omega_r t} - 1) - \frac{1}{2}(e^{i2m\omega_r t} - 1)\right] \\ -\frac{1}{32m\omega_r}\left[\frac{1}{i2m\omega_r}(e^{i2m\omega_r t} - 1) - t\right]\end{array}\right\} I_Y$$

$$(110)$$

which is simplified after calculation

$$F_{2,0}^Y(t) = \frac{1}{2}\left(\sum_{m=-2}^{2}\omega_I^{(m)}\right)^2 \left(\frac{1}{m^2\omega_r^2 i}\right)\left\{\frac{1}{48}e^{i3m\omega_r t} + \frac{1}{64}(-23 + i34m\omega_r t)e^{i2m\omega_r t} - \frac{1}{16}e^{im\omega_r t} + \frac{77}{192}\right\} I_Y \quad (111)$$

We can calculate the expression at $\tau_C = \frac{2\pi}{\omega_r}$,

$$F_{2,0}^Y(\tau_C) = \frac{1}{2}\left(\Sigma_{m=-2}^2 \omega_I^{(m)}\right)^2 \left(\frac{1}{m^2\omega_r^2 i}\right)\left\{\frac{1}{48}e^{i6m\pi} + \frac{1}{64}(-23 + i68m\pi)e^{i4m\pi} - \frac{1}{16}e^{i2m\pi} + \frac{77}{192}\right\}I_Y \tag{112}$$

and

$$\frac{iF_{2,0}^Y(\tau_C)}{\tau_C} = \frac{1}{2}\left(\Sigma_{m=-2}^2 \omega_I^{(m)}\right)^2 \left(\frac{1}{2\pi m^2\omega_r}\right)\left\{\frac{1}{48}e^{i6m\pi} + \frac{1}{64}(-23 + i68m\pi)e^{i4m\pi} - \frac{1}{16}e^{i2m\pi} + \frac{77}{192}\right\}I_Y = A_Y I_Y \tag{113}$$

$$F_{2,0}^Z(t) = -\frac{1}{2}\left(\sum_{m=-2}^{2} \omega_I^{(m)}\right)^2 \begin{Bmatrix} \frac{-2}{16im^2\omega_r^2}\left[\frac{1}{7}\left(e^{i\frac{7}{2}m\omega_r t} - 1\right) - \frac{1}{3}\left(e^{i\frac{3}{2}m\omega_r t} - 1\right)\right] + \frac{1}{8im^2\omega_r^2} * \\ *\begin{bmatrix}\frac{1}{5}\left(e^{i\frac{5}{2}m\omega_r t} - 1\right)\\ +\left(e^{i\frac{1}{2}m\omega_r t} - 1\right)\end{bmatrix} + \frac{1}{18im^2\omega_r^2}\left[1 + \left(i\frac{3}{2}m\omega_r t - 1\right)e^{i\frac{3}{2}m\omega_r t}\right] \\ -\frac{1}{2im^2\omega_r^2}\left[\left(\frac{1}{2}m\omega_r ti - 1\right)e^{i\frac{m}{2}\omega_r t} + 1\right] \\ +\frac{1}{12m\omega_r}\left[\frac{2}{i7m\omega_r}\left(e^{i\frac{7}{2}m\omega_r t} - 1\right) - \frac{1}{i2m\omega_r}\left(e^{i2m\omega_r t} - 1\right)\right] \\ -\frac{1}{4im^2\omega_r^2}\left[\frac{2}{5}(e^{i\frac{5}{2}m\omega_r t} - 1)\right. \\ \left. -\frac{1}{2}(e^{i2m\omega_r t} - 1)\right] \\ -\frac{1}{12m\omega_r}\left[\frac{2}{i3m\omega_r}\left(e^{i\frac{3}{2}m\omega_r t} - 1\right) - t\right] \\ +\frac{1}{4m\omega_r}\left[\frac{2}{im\omega_r}\left(e^{i\frac{m}{2}\omega_r t} - 1\right) - t\right] \end{Bmatrix} I_Z \tag{114}$$

which is simplified after calculation,

$$F_{2,0}^Z(t) = \frac{1}{2}\left(\Sigma_{m=-2}^2 \omega_I^{(m)}\right)^2 \left(\frac{1}{m^2\omega_r^2 i}\right)\left\{\frac{1}{168}e^{i\frac{7}{2}m\omega_r t} - \frac{3}{40}e^{i\frac{5}{2}m\omega_r t} + \frac{1}{12}\left(-\frac{5}{6} + im\omega_r t\right)e^{i\frac{3}{2}m\omega_r t} + \left(\frac{9}{8} - \frac{1}{4}im\omega_r t\right)e^{i\frac{m}{2}\omega_r t} + \frac{1}{12}e^{i2m\omega_r t} - \frac{1}{6}im\omega_r t - \frac{2359}{2205}\right\}I_Z \tag{115}$$

We can calculate the expression at $\tau_C = \frac{2\pi}{\omega_r}$,

$$F_{2,0}^Z(\tau_C) = \frac{1}{2}\left(\sum_{m=-2}^{2} \omega_I^{(m)}\right)^2 \left(\frac{1}{m^2\omega_r^2 i}\right)\left\{\frac{1}{168}e^{i7m\pi} - \frac{3}{40}e^{i5m\pi} + \frac{1}{12}\left(-\frac{5}{6} + i2m\pi\right)e^{i3m\pi} + \left(\frac{9}{8} - \frac{1}{4}i2m\pi\right)e^{im\pi} + \frac{1}{12}e^{i4m\pi} - \frac{1}{6}im2\pi - \frac{2359}{2205}\right\} I_Z \quad (116)$$

and

$$\frac{iF_{2,0}^Z(\tau_C)}{\tau_C} = \frac{1}{2}\left(\sum_{m=-2}^{2} \omega_I^{(m)}\right)^2 \left(\frac{1}{2\pi m^2\omega_r}\right)\left\{\frac{1}{168}e^{i7m\pi} - \frac{3}{40}e^{i5m\pi} + \frac{1}{12}\left(-\frac{5}{6} + i2m\pi\right)e^{i3m\pi} + \left(\frac{9}{8} - \frac{1}{4}i2m\pi\right)e^{im\pi} + \frac{1}{12}e^{i4m\pi} - \frac{1}{6}im2\pi - \frac{2359}{2205}\right\} I_Z = A_Z I_Z \quad (117)$$

which leads to the calculation of the expression of the first-order average Hamiltonian for the Fer expansion,

$$\overline{H_{Fer}^{(1,0)}} = \frac{iF_{2,0}(\tau_C)}{\tau_C} = A_X I_X + A_Y I_Y + A_Z I_Z + \sum S^{(10)}. \quad (118)$$

where $\sum S^{(10)}$ are the sum of the equivalent terms of spin I for spin S, and the functions, $A_X$, $A_Y$, and $A_Z$ are calculated and results are given above in the Eqs. (109), (113), and (117), respectively. The propagator derived from the FE can be written as

$$U_{Fer}(\tau_C) \approx \exp\left\{-i\tau_C\left(\overline{H_{Fer}^{(0)}} + \overline{H_{Fer}^{(1,0)}} + \cdots\right)\right\} \quad (119)$$

where $\overline{H_{Fer}^{(0)}}$, $\overline{H_{Fer}^{(1,0)}}$, ... etc correspond to the zero[th], 1[st] ... etc... average Hamiltonian orders for the Fer expansion, respectively.

## VII. Comparison and Discussion

The Eqs. (47), (48), (93) and (99) are the evaluation of the first-order contributions for the FME and the zero-order contributions for the FE, respectively. Both approaches are identical at their respective first level. These results show that at the lowest order of expansion of the FME and FE, both approaches converge to the same results during the TOFU pulse sequence radiation experiment. The FME and FE approaches were developed for the improvement of analytic methods for studying quantum systems driven by time-dependent Hamiltonians in theoretical physics and found their applicability in wide range of problems that involve coherent control and manipulation of quantum systems[80]. Initially, both methods (FME and FE) starts with the development of an alternate formula to evaluate the time-propagator ($U(t, 0)$) given by,

$$U(t, 0) = \hat{T} \exp\left(-i \int_0^t H(t') dt'\right) \quad (120)$$

where $\hat{T}$ represents the time-ordering operator. However, the operator $\hat{T}$ prevent the direct integration of the operators in the exponent appearing in the above $U(t,0)$ Eq. (120). Furthermore, the calculation of the propagator $U(t,0)$ in systems in control by time-dependent Hamiltonians is further intensified by the non-commutativity of the Hamiltonians at different times. Both analytical methods (FME and FE) are expressed as exponential operator representation (Eqs. (11), (13), and (24)), which explain the reason why the two approaches are identical at their respective first level. In addition, one key property that needs to be ensured for both approaches is the conservation of unitarity at any finite order of series applicable to any time dependent Hamiltonian, which satisfies the physical behavior such as the non-violation of norm conservation.

The Eqs. (70-1), (82), and (118) are the evaluation of the second-order contributions for the FME and the first-order contribution for the FE, respectively. Both approaches are not identical at their respective second level of contribution. The propagator derived from the FME (Eq. (83)) is different to the propagator derived from the FE (Eq. (119)). The results of the FME (Eq. (83)) compared to results of the FE (Eq. (119)) show that the FME approach converges more faster than the FE during the TOFU pulse sequence radiation experiment. The spin dynamics evolution is much more complexed during the application of FE compared to FME. The low performance of the FE during this scheme can be linked or associated to the complexity of spin dynamics during its evolution. This outlines the serious FE limitations with regard to the description of time-evolution of quantum systems. Such as derived in section IV, depending on the form of the initial density operator and the detection operator, the final form of the signal expression derived from the two formulations could differ as showed in the Eqs. (32-b) and (33-b). In the time-propagators based on Left running expansion, the $F_n(t)$ operators (of higher order 'n') act initially on the initial density operator, while in the Right running expansion, the $F_n(t)$ operators (of lower order 'n') act initially on the initial density operator. This work supports is in agreement with the recent work of Ramachandran[80] that also highlighted the over dependence on the commutator relations between the operators ($[F_n(t), \rho(0)], [F_n(t), \hat{D}]$), which limits the utility of the FE approach in time-evolution studies. It is noteworthy that for the most part, the current literature stresses more effort on the derivation of the $F_n(t)$ operators than the accuracy of the FE method. Based on the approaches revisited in section IV and the results of the Eqs. (118) and (119) for FE compared to the Eqs. (70-1) and (70-2) for FME, we can conclude that, the FE scheme works only in special cases where the conditions $[F_n(t), \rho(0)] \neq 0 \text{ and } [F_n(t), \hat{D}] \neq 0$ are both completed.

## VIII. Conclusion

It is worth stressing that this is the first attempt to apply the FME and FE approaches to investigate the TOFU radiation experiment. We have revisited the accuracy of the FE scheme and the FME approach as complementary methods to the well-established methods in solid-state NMR, namely, the average Hamiltonian theory[61-63] and the Floquet theory[71-73]. We compared the FME and FE approaches based on the quantification of the effective Hamiltonians and propagators, and we found that the FME approach is more appropriate to describe the dynamics of spin system during the TOFU pulse sequence radiation experiment compared to the FE approach. This manuscript brings the relevance or the subtle points of the two expansion schemes and gives new insights into the analysis of the TOFU scheme. It is also important to highlight that many aspects of the TOFU pulse sequence have still not been worked out. The comparison between the FME and FE give us an indication on the performance of the two approaches as well as the relevance of the cross-term effects. Our analysis might suggest the possibility to study the resonance condition, what limits the resolution in each case of the recoupling, how the resolution can be improved, and issues related to all these points. This pilot study could show more that the FME is a promising approach to the study of more complex pulse sequences. Future theoretical works are certainly feasible and we hope to approach also other important pertinent point such as to do an actual line shape calculation due to the TOFU experiment with certain orders of the schemes that will bring out the relevance of the argument. We showed in this article that elegant integral calculations are possible with the two approaches and that the differences between FME and FE are consistent and the similar calculations could be applied to other sequences as well. We believe that all the expressions derived can be employed for numerical time evolution for a broad class of time dependent Hamiltonians and thus can be utilized for calculating observables in time resolved spectroscopy, quantum control, and quantum sensing as well as for the open system quantum dynamics when the reduced system density operator is solved in the interaction picture with respect to the zeroth order system Hamiltonian.

**Acknowledgments**

The author thanks Dr Thibault Charpentier for useful comments. He acknowledges the support from the CUNY Research Foundation for the PSC-CUNY Research Award (TRADB-54-75, Award # 66377-00 54). The contents of this paper are solely the responsibility of the author and do not represent the official views of the NIH.

**Conflict of Interest**

The authors confirm that this article content has no conflict of interest.

**Appendix**

**1.**

$$cos^2(Ct) = \frac{1}{2} + \frac{1}{4}\exp(i2Ct) + \frac{1}{4}\exp(-i2Ct) \tag{A.1}$$

$$sin(Ct) = \frac{1}{2i}(e^{iCt} - e^{-iCt}) \tag{A.2}$$

$$sin(2Ct) = \frac{1}{2i}(e^{i2Ct} - e^{-i2Ct}) \tag{A.3}$$

**2.**

$$F_2^\sigma = \left[\frac{1}{2}\sum_{m=-2}^{2}\omega_I^{(m)}\left(\frac{1}{6m\omega_r}\right) + \frac{1}{8}\sum_{m=-2}^{2}\omega_I^{(m)}\left(\frac{2}{3m\omega_r}\right)\right]I_X$$

$$+\left[\frac{-1}{64m\omega_r}\sum_{m=-2}^{2}\omega_I^{(m)}\right.$$
$$-\sum_{m=-2}^{2}\omega_I^{(m)}\left(\frac{1}{64m\omega_r}\right)$$
$$\left.-\sum_{m=-2}^{2}\omega_I^{(m)}\left(\frac{1}{64m\omega_r}\right)-\sum_{m=-2}^{2}\omega_I^{(m)}\left(\frac{5}{64m\omega_r}\right)\right]I_Y$$
$$+\left[0-\sum_{m=-2}^{2}\omega_I^{(m)}\left(\frac{1}{12im\omega_r}\right)-\sum_{m=-2}^{2}\omega_I^{(m)}\left(\frac{1}{12im\omega_r}\right)\right]I_Z+\sum S^{(3)}$$
$$=\sum_{m=-2}^{2}\omega_I^{(m)}\left(\frac{1}{2m\omega_r}\right)\left[\frac{1}{3}I_X-\frac{1}{4}I_Y-\frac{1}{3i}I_Z\right]+\sum S^{(3)}$$